\documentclass[11pt,twoside]{article}

\usepackage{asp2006}
\usepackage{epsf}
\usepackage{psfig}
\usepackage{lscape}

\begin{document}
\title{Conference Summary}   
\author{Richard S Ellis}   
\affil{California Institute of Technology}    

\begin{abstract} 
This first Subaru international conference has highlighted the remarkably diverse
and significant contributions made using the 8.2m Subaru telescope by both 
Japanese astronomers and the international community. As such, it serves as a
satisfying tribute to the pioneering efforts of Professors Keiichi Kodaira and Sadanori 
Okamura whose insight and dedication is richly rewarded. Here I try to summarize the 
recent impact of wide field science in extragalactic astronomy and cosmology
and take a look forward to the key questions we will address in the near future. 
\end{abstract}


\section{The Unique Role of Subaru}   

It's a pleasure to summarize the first Subaru international conference. The conference
has been striking in many ways. Before I arrived in Hayama, I was impressed
by one of the two conference posters: the woodblock print (Ukiyo-e) `The Great Wave
off Kanagawa" by the artist Katsushika Hokusai (1760-1849). This portrays desperate
fishermen struggling with a tsunami against the backdrop of Mt Fuji. I thought this seemed an 
alarming image to use to encourage overseas visitors to come to the meeting, particularly given 
the proximity of the conference venue to the ocean and the location depicted! Fortunately,
I'm glad to report our views of Mt Fuji and the ocean have been peaceful and serene throughout
our stay.

The second striking thing about the meeting is the cause for celebrating the achievements
of the Subaru telescope. Although we've heard panoramic results from many observatories
and facilities, it's clear much of the focus has been with Subaru. We're congratulating
Professor Keiichi Kodaira for his leadership in bringing this remarkable facility to fruition,
and Professor Sadanori Okamura and his colleagues for their insight in developing the
remarkable prime focus camera, Suprime-Cam. Subaru remains the only fully-steerable
8 meter class telescope with a prime focus imager; we've seen ample evidence
of the wisdom of this decision. With further instruments such as
MOIRCS, FMOS and hopefully eventually WFMOS, the Subaru telescope is destined to remain
an unique facility in wide-field astronomy.

In an interesting presentation, Iye (2007) examined a recent article by Grothkopf
et al (2007) which analyzed the Hirsch (2005) $h$-index\footnote{The $h$-index is
the number $n$ of refereed articles each of which has more than $n$ citations} of the 
four major 8-10m telescopes. Subaru ranks third in this analysis (behind Keck and 
the VLT) with a $h$-index of 40. However, as citations are cumulative, older facilities 
naturally garner more citations. The $m$-index takes this into account by dividing
the number of citations by the years of routine operation. Grothkopf et al find that
when this is done, the newer VLT is having a comparable impact to Keck, both
have $m\simeq$10, but  Iye was dismayed to find Subaru remains third with $m\simeq$6.

Not one to be defeated easily, Iye invented the $i$-index which is the $m$-index
divided by the number of telescopes in the observatory. As the VLT comprises 4 
independent telescopes, and Keck and Gemini comprise two each, Subaru is 
clearly disadvantaged by this factor. When this correction is made, as a single 
telescope, Iye was pleased to find Subaru leads the pack!

Even if citations are never a fully-accurate guide of the discovery rate of
a telescope, I think we'd all agree that Subaru has, through its unique wide 
field capabilities, contributed enormously to the growth and international standing 
of Japanese astronomy.

\section{Cosmic Dawn: Onward Into the Dark Ages}

One of the most active areas discussed at the meeting concerns the
study of the earliest galaxies observed in the redshift interval 4$<z<$10, corresponding
to the first 2 Gigayears of cosmic history. Although originally the province of 
Hubble Space Telescope through public deep imaging campaigns locating
UV continuum sources (Lyman break galaxies, LBGs), ground-based programs 
largely led by imaging with Subaru are now just as prominent. These offer complementary 
datasets both in terms of increased areal coverage, and via the location and study of 
Lyman $\alpha$ emitters (LAEs) through narrow band imaging. The results we have seen 
are striking, both in terms of the rapid observational progress and the puzzles they 
raise.

At this meeting we have witnessed the contrast in the decline in star formation beyond 
$z\simeq$3 observed in the LBGs and LAEs (contributions by Kashikawa, Bouwens, 
Capak, Ouchi and Ohta). A strikingly different pattern is seen (Figure 1). There is a 
marked, possibly luminosity-dependent, evolution seen in the LBG population, but 
almost no evolution is seen in the LAEs. Various comparative studies have suggested 
LAEs are generally less massive and younger than their equivalent LBGs, but 
why the Ly$\alpha$ luminosity function should conspire to be unchanging over 
3$<z<$6 when there is such a significant change in the LBG population is 
particularly intriguing.

The constancy of the Ly$\alpha$ luminosity function for LAEs over 3$<z<$5.7 
serves to emphasize the significance of the remarkable drop in the abundance
of luminous emitters over 5.7$<z<$6.5 as presented by Kashikawa (Figure 2).
This is the best available data at the present time and has been used
to imply an abrupt increase in the neutral fraction beyond $z\simeq$6 which could be as 
high as $x_{HI}\simeq$0.4 at $z\simeq$6.5. This contrasts with independent
estimates of $x_{HI}$, e.g. along the line of sight to the only equivalent high 
redshift gamma ray burst, GRB050904 (Kawai) and perhaps the presence of
Ly$\alpha$ in higher redshift sources (Iye et al 2006).

At face value Figure 2 suggests that star formation was sufficiently vigorous
at or just before $z\simeq$6 to initiate reionization of the Universe. But, unless 
the relevant physical parameters (escape fraction, slope of the stellar initial 
mass function etc) are extreme, until recently most workers believed the star 
formation density observed in LBGs was insufficient (Bunker et al 2004). So the 
result from Kashikawa et al (2006) is puzzling but still potentially very significant. 
Enlarging the survey area over which the LAE luminosity function can be
determined (particularly at $z\simeq$6.5) if thus a high priority.

\vspace{0.1in}
\centerline{\hbox{\psfig{file=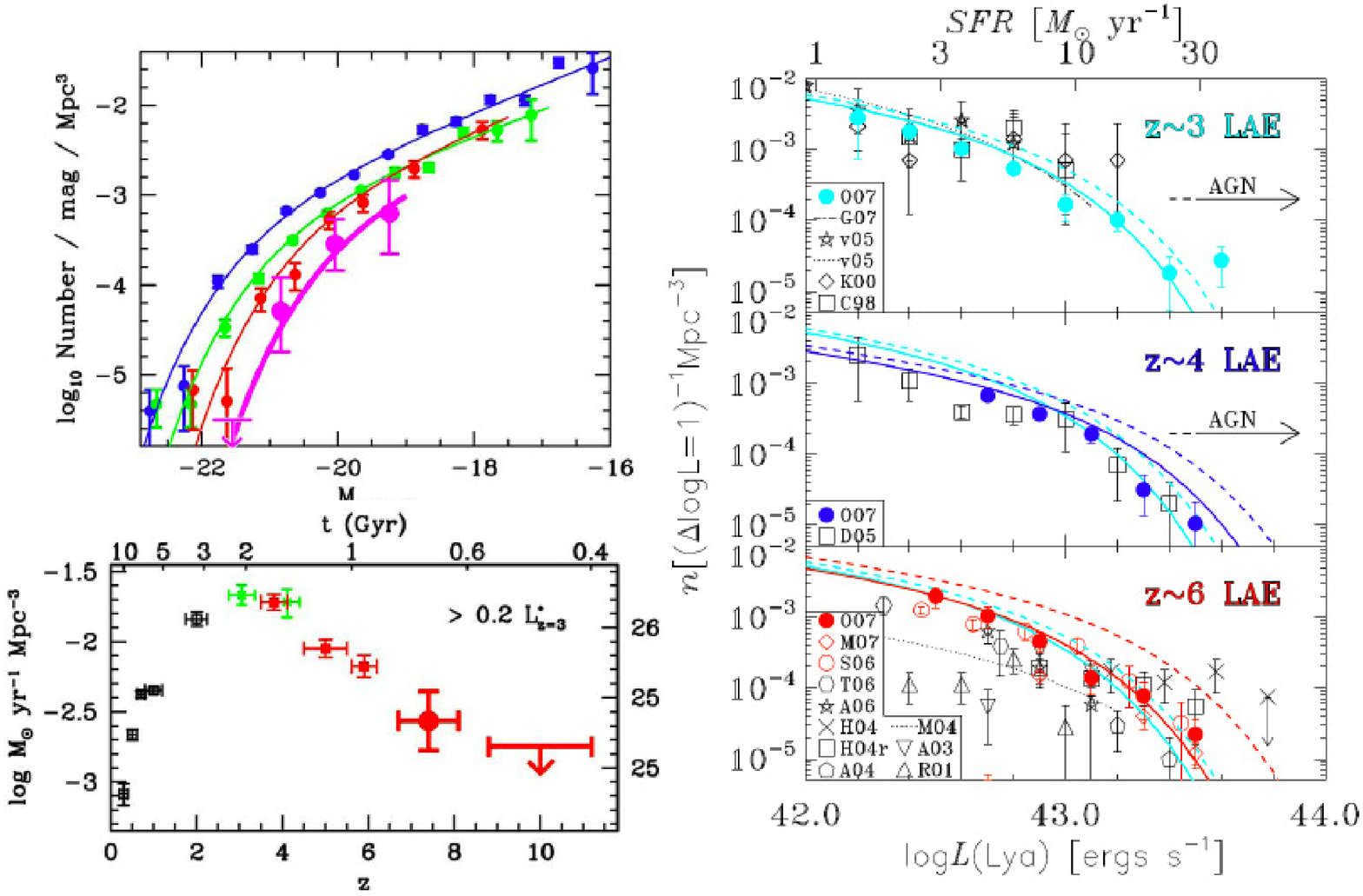,width=5.2in,angle=0}}}

\noindent{\small \textsf{\bf Figure 1: } Contrasting the redshift evolution
of Lyman break and Lyman alpha emitting galaxies over 3$<z<$6
from the surveys of Bouwens (left), Kashikawa and Ouchi (right)
and their respective collaborators (Bouwens et al 2007, Ouchi et al 2008).}
\smallskip

Indirect evidence using IRAC that supports much earlier star formation activity 
arises from the presence of massive galaxies with established stellar populations at $z\simeq$5-6 
(Eyles et al 2006, Stark et al 2007). These workers located a few
spectroscopically-confirmed galaxies with stellar masses $\simeq$ 2-12 $\times$
10$^{10} M_{\odot}$, and Balmer breaks in their energy distributions, implying 
the bulk of the responsible star formation in these sources occurred in 
the redshift interval 7$<z<$15. 

It's a big step from locating a few massive galaxies with established stellar
populations to measuring the comoving density of stellar mass at $z\simeq$5-6
necessary to provide a global integral constraint on earlier activity. In doing
so we are limited both by spectroscopic incompleteness, uncertainties
in the derived stellar masses and IRAC confusion. McLure showed us how 
stacking LBGs in the ongoing UKIDSS survey can provide new constraints
improving significantly over the cosmic variance limited results in the
deep HST fields.

\vspace{0.1in}
\centerline{\hbox{\psfig{file=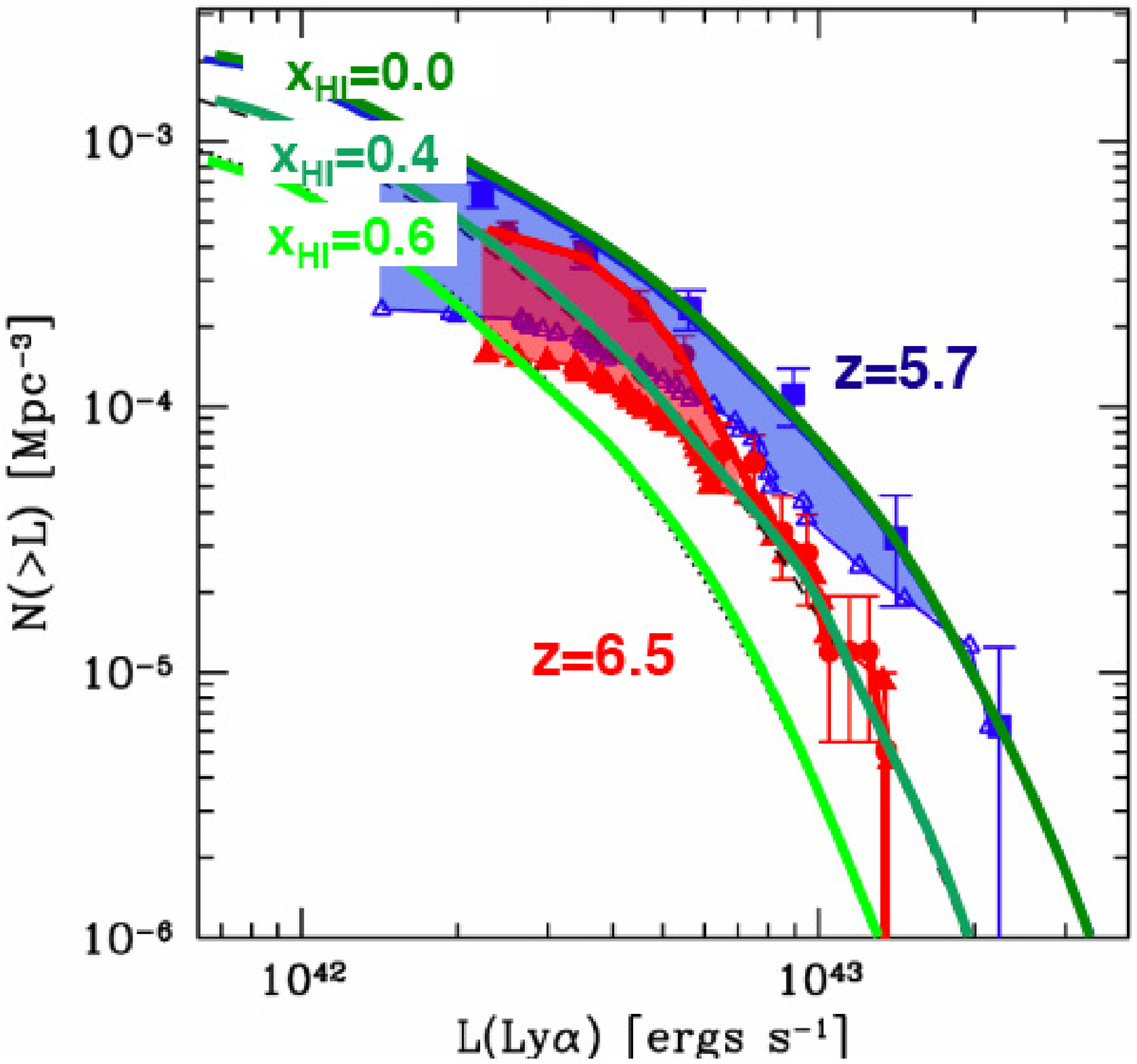,width=3.0in,angle=0}}}

\noindent{\small \textsf{\bf Figure 2: } The remarkable
discontinuity in the abundance of LAEs over 150 Myr 
from $z$=6.5 to $z$=5.7, as presented by Kashikawa et al (2006).
Lines represent the estimated effect of a decreasing neutral fraction
from $z$=6.5 to 5.7.}
\smallskip

My belief is that the evidence for an active era of star formation in the interval
$7<z<$10 is fairly strong. In addition to the presence of old stars and the
integrated mass density, further support comes from the presence
of CIV absorption in the spectra of $z\simeq$6 QSOs observed with the VLT
(Ryan-Weber et al 2006). Taking all of the foregoing into account, it is perhaps 
surprising, therefore, that we have such difficulty finding luminous precursors at higher
redshift. A possible explanation is that the early sources were obscured, or 
perhaps many are sub-luminous systems which subsequently merge. Whatever 
the explanation, searching for $z>$7 sources is clearly a well-motivated quest.

Ouchi, Tokoku and Stark described exciting attempts to locate candidate star-forming
galaxies in this uncharted region. This is highly challenging work in many respects:
demonstrating convincing candidates, securing spectroscopic confirmation and
removing foreground contaminants. It is unclear what surface density of star
forming sources we expect, so we should not discount surprises or expect
continuity in the decline in abundance with redshift in sub-luminous sources
probed by strong lensing. Rauch et al (2007) demonstrated nicely the unexpected
upturn in the surface density of faint emitters at lower redshift. In these 
pilot programs we are really witnessing the first glimpse of the era that
will be explored with next generation facilities such as JWST and TMT.

\section{Structure Formation: What Evolves into What?}

Panoramic data also enables us to measure the degree of clustering in
faint sources which allows us to estimate, at a given redshift, the bias parameter
for a population with respect to the underlying dark matter. Such data gives us 
valuable insight into how to connect one population into another, for example
at lower redshift (talks by Coil, Wechsler and others). Likewise, by studying a 
few selected areas with a variety of facilities, the demographics of various populations 
can be understood in the same way that the density of stars at a location on 
the Hertzsprung-Russell diagram gives insight into the time stars spend during 
various evolutionary stages.

Improved photometric redshifts will considerably sharpen all of these associations
and clustering studies. Scoville and Ilbert showed how the provision of 34 filter photometry
in the COSMOS field has enabled an unprecedented photometric redshift precision 
of $\sigma_z\,/\,(1 + z)$ = 0.007 for a sample of 6000 galaxies with $i<$22.5 
studied with the VLT, Keck and Magellan over 0$<z<$1.5 and $\sigma_z\,/\,(1 + z)$ = 0.016
for 550 XMM-selected AGN. One can imagine further improvements in photo-z
codes as the overlap with spectroscopic samples increase. And although not all
areas will have as rich a dataset as the COSMOS field, conceivable we can use it
as a training set for wider less well-sampled areas.

Shimasaku showed us that an astonishing 17,000 Lyman break galaxies have now
been located beyond $z\simeq$4 and the overall bias-redshift relation can be traced
for various populations over 0$<z<$4 (Figure 3). Connections made 
between selected populations seen at different epochs can only be tentative 
because of the vagaries of luminosity selection and the unknown duty cycle
of star formation activity. Additional information, for example stellar timescales 
revealed by spectroscopic data, may be needed. One senses we still have some 
way to go understand the general route from high redshift Lyman break, 
sub-mm and distant red galaxies through to the local population of massive galaxies,
but the progress is good.

\vspace{0.1in}
\centerline{\hbox{\psfig{file=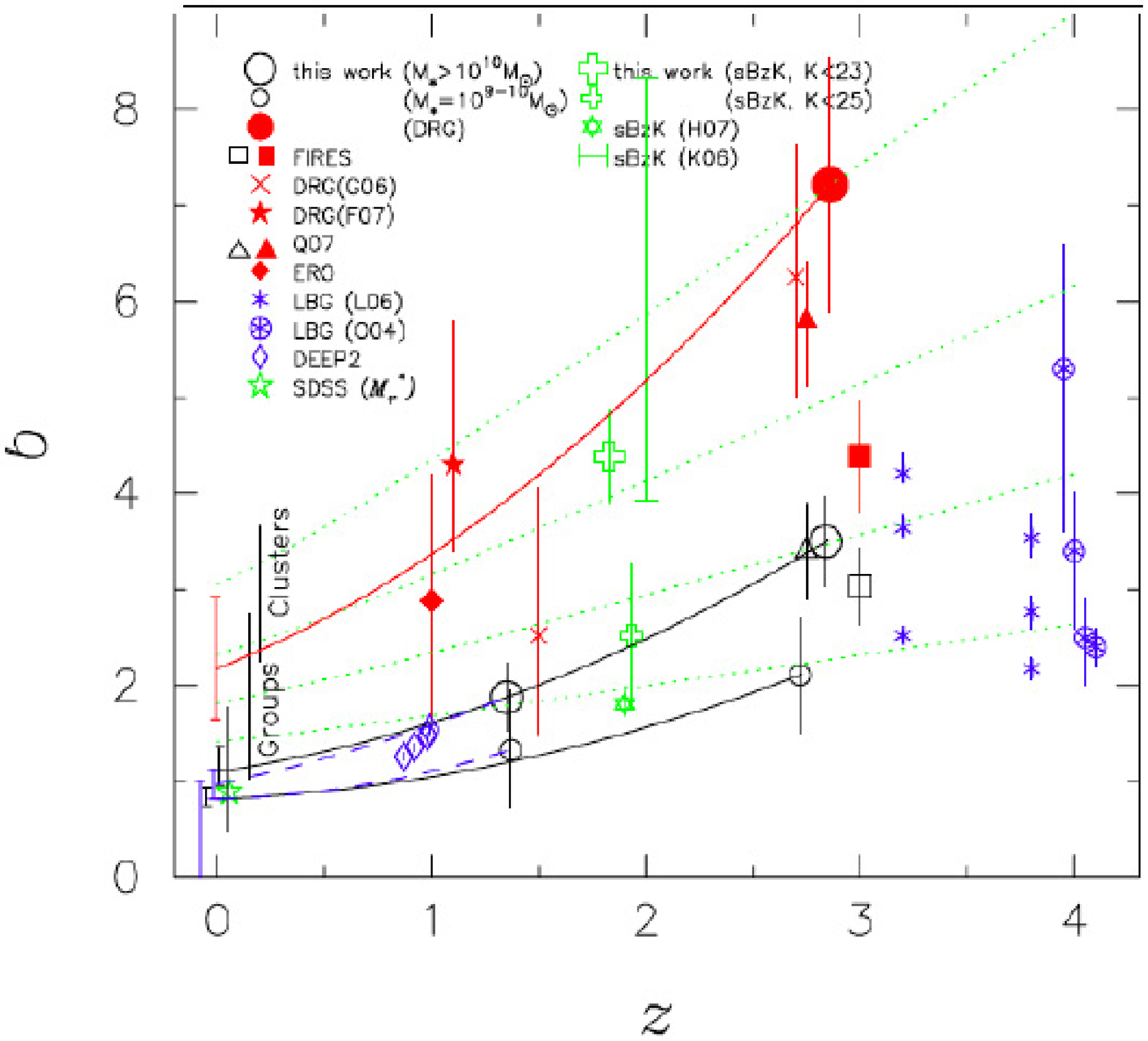,width=4.0in,angle=0}}}

\noindent{\small \textsf{\bf Figure 3: } A summary of the bias parameter $b$
determined from the clustering of various types of galaxies selected at a given 
redshift from the compilation of Hayashi et al (2007). Such data allows the
associated halo mass to be determined and hence can, in principle, guide 
how one population might evolve into another at lower redshift.}
\smallskip

A more tangible goal was demonstrated in Selected Area SSA22. Here, Yamada, 
Nakamura, Kohno and colleagues have surveyed a field extending almost over 
200 $\times$ 120 comoving Mpc with a variety of facilities locating Lyman $\alpha$ `blobs', 
high equivalent width emitters and dusty star forming sources. Each type of source 
might be considered a transient phase in the development of a later, more mature, 
star-forming or quiescent galaxy. In addition to studying the environmental densities 
where such sources occur, the relative abundances and cross-linking of these interesting 
populations is giving us a first glimpse at their demographics and inter-relationship. 
This is exciting work which can easily be extended over wider fields.

Scoville also showed us how weak lensing now defines directly the relative distribution
of dark matter (DM) against which various baryonic tracers can be compared; the agreement
is striking (Massey et al 2007). Although the fidelity of the projected DM map using
the Hubble's ACS is likely better than that from ground-based imagers,
the study of these correlations over larger scales aided with photometric data, will
surely be a major application for HyperSuprime-Cam.\footnote{During the actual meeting, we
learned that the COSMOS DM map was ranked as one of the top 10 science stories
of 2007 by Discover Magazine - alongside global warming in the arctic and the discovery
of proteins in Tyrannosaurus Rex!}

\section{Mass Assembly: Puzzles Remain}

It's clear that the widespread availability of multi-wavelength data has enabled
a bewildering array of surveys motivated by the question of measuring the 
time-dependent star formation rate and mass assembly. 

We heard observational talks on this topic from Fontana, Dickinson, Ilbert, Taylor, 
Conselice, Cirasuolo and Egami. Theoretical interpretation of these results in
the context of popular semi-analytical models were presented by Bower and
Somerville. 

The observational results did not, in my opinion, always agree, particularly for the 
stellar mass function $\Phi(M,z)$ and its dependence with galaxy type. And I
think the theorists were somewhat selective in their choice of observational surveys
with which they compared their predictions. Several speakers asserted definitive 
conclusions, for example, regarding the evolution of the mass function for quiescent 
and star-forming galaxies over 0$<z<$2 yet, to me,  there seems quite a diversity of 
observational conclusions on this point (see discussion in Stringer et al 2008).

It seems unfortunate we have so many groups working on this topic without some 
form of rigorous comparison. Conceivably we need some form of blind test for the
derivation of stellar masses from photometric data, and an impartial independent 
reviewer of the various survey results is sorely needed. 

The following technical problems occur to me in listening to the various observational
results:

\begin{itemize}

\item{} Cosmic variance is a major factor in all of the surveys undertaken so far.
Error bars defined from the counting statistics is a poor basis for estimating the 
significance of any trend observed (c.f. recent paper by Stringer et al 2008).

\item{} Beyond $z\simeq$0.7, reliable stellar masses require high quality
near-infrared data; the contribution from established stellar populations is 
otherwise poorly-estimated. A surprising fraction of the survey results are
based on stellar masses derived from optical data alone.

\item{} A good precision in photometric redshift does not imply an equivalent
precision in the derived stellar mass, particularly if both are derived from the
same photometric data. The cross-talk in deriving both from samples without 
spectroscopy can lead to major uncertainties (see Fig.~3 of Bundy et al 2005).

\item{} Understanding the impact of completeness is important given the 
(usually) optically-limited spectroscopic or photometric samples.

\end{itemize}

In the huge amount of information presented during both the observational and
theoretical talks in this area, I've selected three scientific issues for further discussion.

The first topic is {\em cosmic downsizing} - a result pretty well all observers
have demonstrated from their evolving mass functions whereby
massive galaxies complete their star-formation earlier than those in lower
mass galaxies\footnote{It's amusing to me how theorists were initially
dismissive of the observers' claims for `anti-hierarchical' evolution yet are
now telling the observers how important feedback is.}
Bower, Somerville and P. Hopkins each discussed the
possible physical origin of this counter-intuitive behavior. Bower contrasted
`radio mode' feedback (Croton et al 2006, Bower et al 2006, Okamoto
et al 2007) with `quasar mode' feedback (Granato et al 2004, Springel et
al 2005) in the context of reproducing the present-day luminosity function
of galaxies, arguing that the shape of the luminosity function is governed 
by the interplay between rapid and hydrostatic cooling. 

But what triggers this feedback? P. Hopkins argued that various mechanisms 
may quench star formation and intermediate redshift data is required to
break the various degeneracies. This can come in the form of 
evolving type-dependenty mass functions (Bundy et al 2006), merger rates 
(Drory et al 2008), AGN fractions (talks by Salim and Renzini) and the radio 
galaxy luminosity function (talk by Sadler).

The availability of the Millenium Simulation (Springel et al 2005) coupled
with semi-analytic algorithms that incorporate various forms of feedback 
means observers are no longer at the mercy of theorists in checking
how well their data can be reproduced. Figure 4 illustrates one such
comparison from Stringer et al (2008) whereby the concept of a {\it threshold
mass} for quenching, first introduced by Bundy et al (2006) is reasonably
well reproduced by models incorporating radio-mode feedback. 

\vspace{0.1in}
\centerline{\hbox{\psfig{file=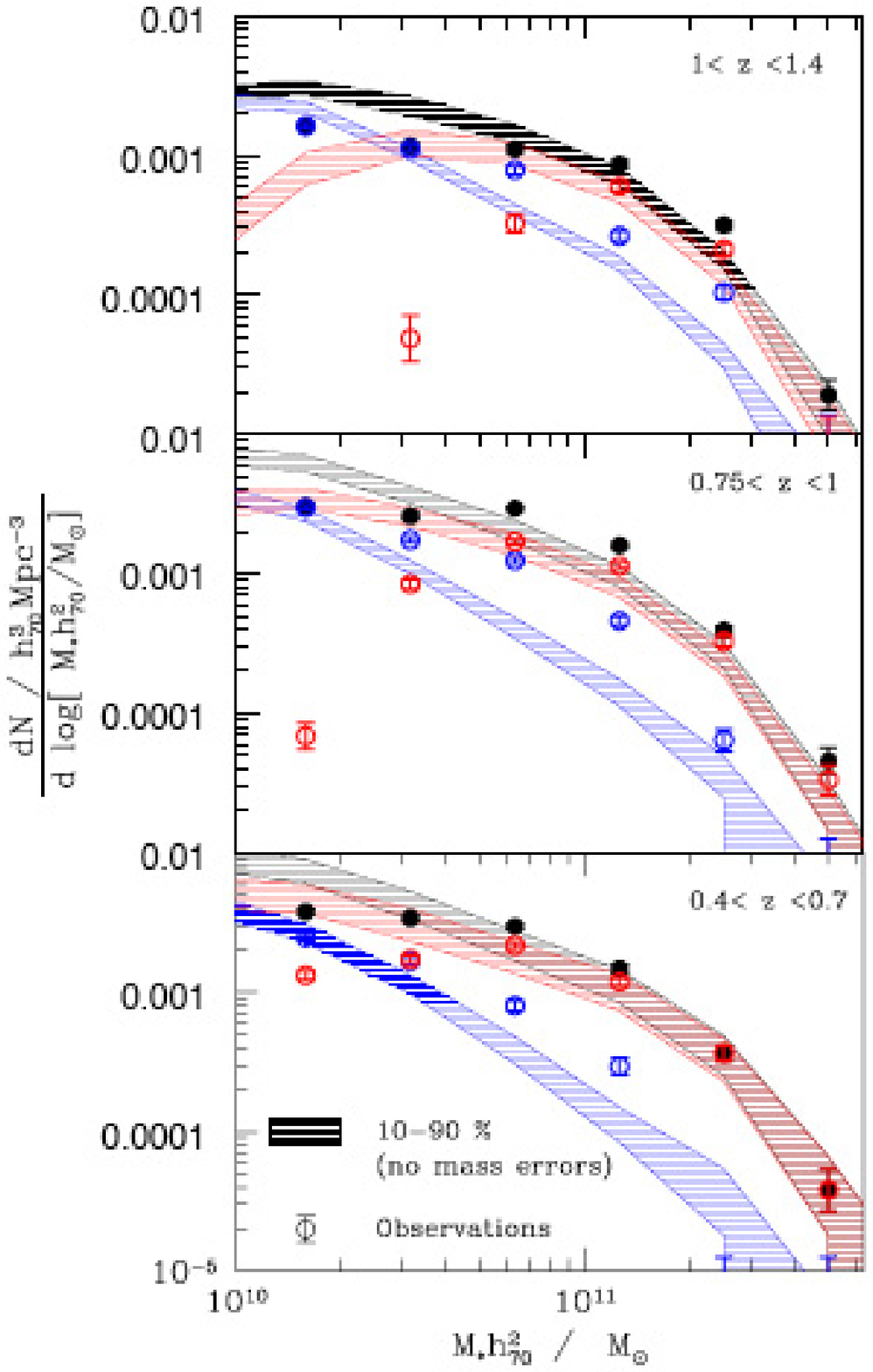,width=4.0in,angle=0}}}

\noindent{\small \textsf{\bf Figure 4:} The evolving stellar mass function
for the total, active and quiescent galaxies from the DEEP2/Palomar 
survey (data points, Bundy et al 2006) compared with the output of the 
GALFORM semi-analytic code incorporating radio-mode feedback in 
suitably-defined light cones in the Millenium Simulation (shaded areas).
The range in the latter illustrates the effect of cosmic variance in the
specific context of the DEEP2/Palomar survey. }
\smallskip

The second interesting highlight was introduced by Mark Dickinson
and relates to the {\em relationship between the star formation rate and
the stellar mass}. Again, he made use of the Millenium Simulation
to show that, at $z\simeq$2, most simulations underpredict the star
formation rate and hence the number of intensely-star forming ultraluminous
infrared galaxies. The tight correlation observed suggests most
star formation at this epoch is not transient hence not driven by
major mergers.

The final highlight is a puzzle introduced by A. Hopkins (see
Wilkins et al 2008) and discussed by Glazebrook, {\it namely our inability to 
reconcile the observed comoving star formation rate density as a function of redshift 
with the observed stellar mass density} (Figure 5). The latter should naturally 
be a time integral of the former and so, at face value, this discrepancy is 
rather alarming. 

What are the possible explanations? Perhaps we have observed
too much star formation at high redshift? This seems unlikely given
so much of it is often obscured (Hopkins \& Beacom 2005). Perhaps 
we have over-estimated the abundance of massive galaxies at 
high redshift? Again, this is an uncomfortable situation since the 
presence of massive galaxies at $z\simeq$2-5 already presents 
some challenges to CDM.

A. Hopkins argued a redshift-dependent initial mass function
may be responsible. This seems like a pretty drastic way out
to me, akin to invoking the cosmological constant when the 
supernova data emerged ! On the other hand some theorists 
will no doubt be delighted with the additional freedom such a 
conjecture brings.  Baugh et al (2005) already have suggested a
top-heavy IMF would fix an embarrasing over-abundance of
sub-mm galaxies whose energetic output is otherwise hard to explain.
My worry is more fundamental; the discrepancy in Figure 5
may be telling us our confidence in the derived quantities
(SFR and stellar masses beyond $z\simeq$1 is unwarranted.
Indeed, perhaps it's suggestive that the discrepancies 
are most apparent at $z>$0.7 where infrared data becomes a
crucial ingredient in deriving stellar masses. 

\vspace{0.1in}
\centerline{\hbox{\psfig{file=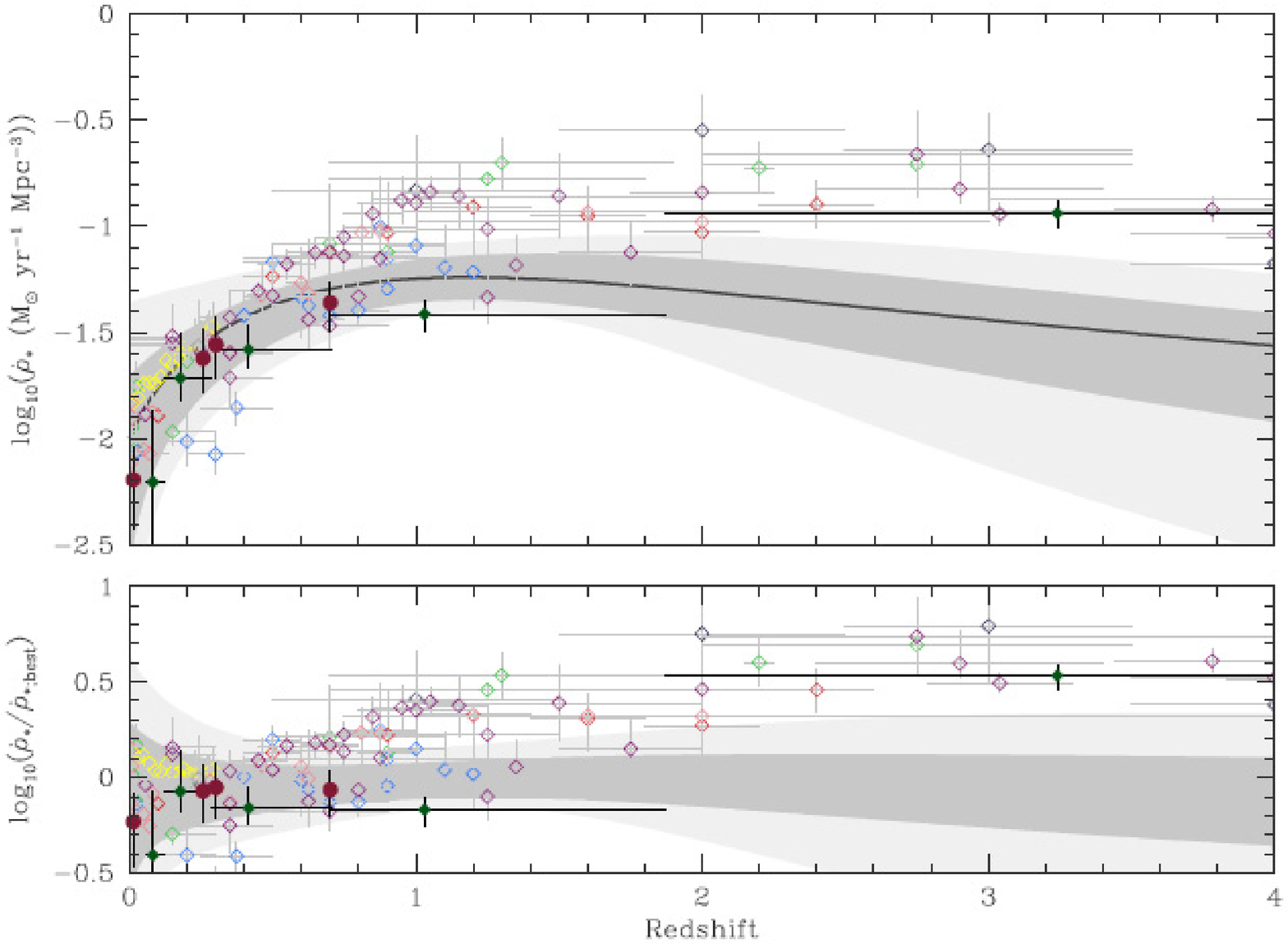,width=4.0in,angle=0}}}

\noindent{\small \textsf{\bf Figure 5: } (Top) The observed comoving density
of star formation as a function of redshift from the compilation
of Wilkins et al (2008) compared with the derivative of the
observed growth of stellar mass (solid line). (Bottom) The 
discrepancy is in the sense of having observed too much star 
formation at high redshift or having underestimated the stellar 
masses of high redshift galaxies. Neither is a particularly
palatable explanation.}
\smallskip

\section{Detailed AO Studies of Individual Sources}

In a digression from the main theme of the conference, I'd like to
highlight the talks by Erb, Lemoine-Busserolle and Akiyama
whose showed the importance of securing resolved imaging and 
spectroscopic data for individual 1$<z<$3 galaxies. The advent
of laser guide star (LGS) assisted observing on several 8-10 meter class
telescopes has meant that all-sky adaptive optics (AO) is finally here!

Erb showed us how Keck LGS-AO with the integral field spectrograph
OSIRIS yields 800 pc resolution at $z\simeq$2-3 with Strehl factors
routinely in the range 25-35\%. In contrast to earlier seeing-limited work 
which necessarily focused on unusually large systems at these
redshifts (e.g. Genzel et al 2005), most appear to be dispersion-dominated
in the sense that a coherent rotation is not seen. Lemoine-Busserolle
is embarking on a large survey of 1.3$<z<$1.6 VVDS galaxies with similar
goals using the SINFONI integral field unit on the ESO VLT. Her first
results are likewise promising.

\vspace{0.1in}
\centerline{\hbox{\psfig{file=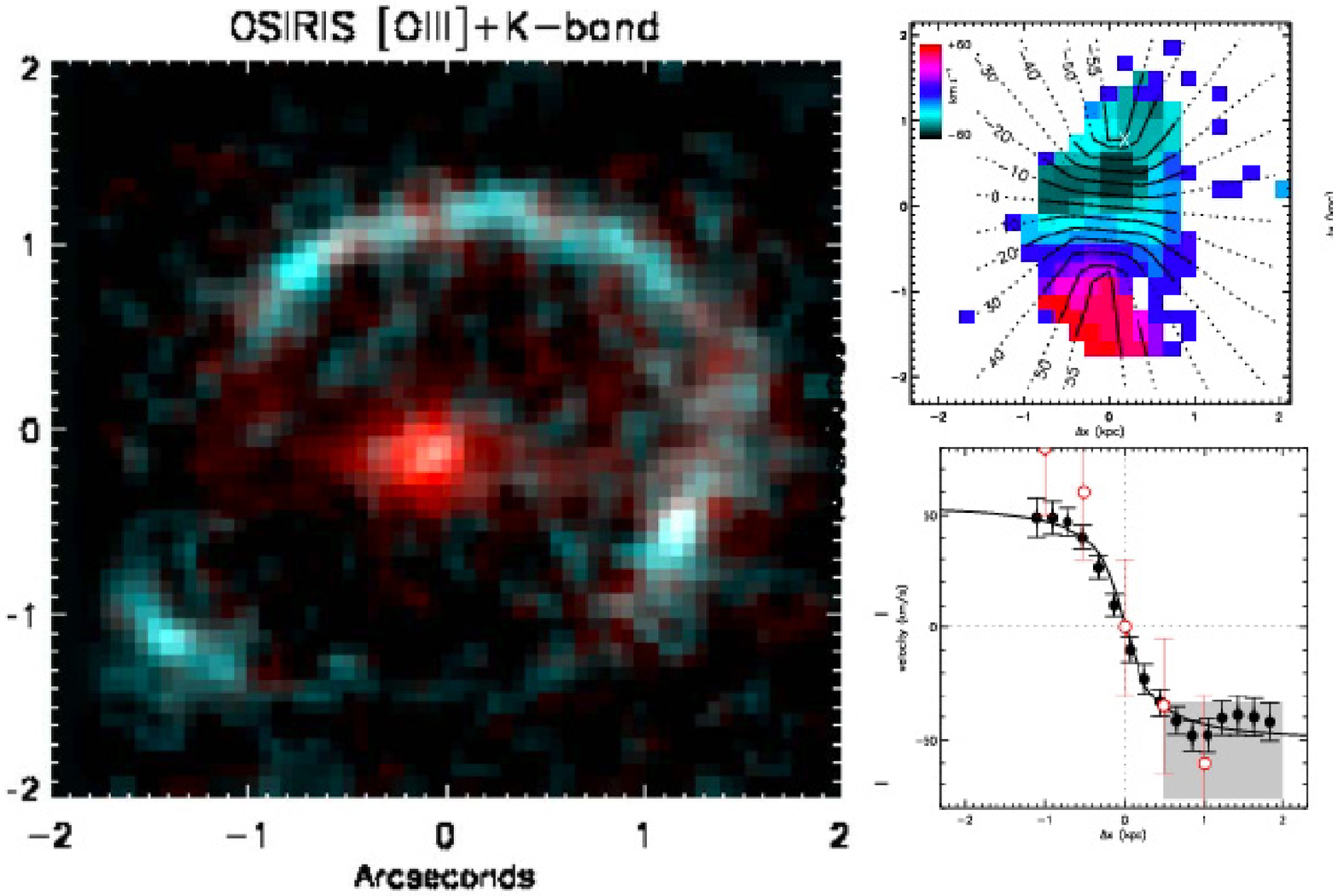,width=4.8in,angle=0}}}

\vspace{0.1in}
\noindent{\small \textsf{\bf Figure 6: } (Left) The "cosmic eye" (Smail
et al 2007), a $z$=3.1 Lyman break galaxies magnified by an areal factor
of $\times$28. The panel combines a HST $V$-band imaging with
Keck LGS AO imaging in redshifted [OIII] emission and $K$-band.
The ground based resolution is 0.13 arcsec. (Right) Source plane velocity field 
and rotation curve derived from the Keck data which, after allowing
for the substantial magnification, achieved an intrinsic resolution of 
80pc (equivalent to 8 milli-arcsec for an unlensed source) (Stark et al 2008).}
\smallskip

Typical $z\simeq$2-3 galaxies are only 1-2 kpc across and thus poorly-sampled 
at 75 milli-arcsec resolution even with the gains of adaptive optics.
However, the angular magnification afforded by gravitational lensing can 
improve this sampling for the small sample of lensed $z\simeq$3 galaxies 
(Swinbank et al 2007, Stark et al 2008, Figure~6). In this case, a resolution of 
better than 100 pc in the source plane can be achieved.

Panoramic imaging surveys such as DES and PanSTARRs and those
planned with HyperSuprime-Cam will generate an enormous candidate
list of such lensed sources, so until the era of TMT, this seems a very
promising route to understand the origin of resolved components (relaxed 
bulges, rotating disks etc) in early galaxies.

\section{Clusters of Galaxies: Evolution and Environmental Effects}

Clusters remains the largest known structures and their abundance and
redshift distribution to early times will provide valuable constraints on
structure formation and cosmology provided we can maintain uniformity
in selection and measure their total masses in a precise, consistent manner.

Clusters are also excellent laboratories for studying the interplay between
dark matter and baryons, for example in merging systems (as nicely shown
in Umetsu's talk), as well as the effect the gravitational potential has
in the environmental transformation of galaxies.

For decades, clusters remained the province of optical and X-ray 
observers. The literature is full of the synergy between measures
of richness and galaxy morphology  (estimated from optical imaging), 
spectroscopic diagnostics of recent star formation (optical spectroscopy) 
and the physical properties of the hot intracluster medium (X-ray imaging
and temperature measures). At this meeting we have seen how infrared
observations from IRAC and MOIRCS have reached fruition and added
significantly to our understanding.

I would highlight the remarkable studies of the emerging red sequence 
presented by Kodama and Tanaka, especially that from MOIRCS $(J-K)$ 
imaging of clusters in the redshift range $2<z<$3 (Figure 7). These results
dovetail beautifully onto the story of how the sequence  subsequently evolves 
differential with luminosity over 0.4$<z<$1 for the EDisCs  (DeLucia)
and.RCS (Yee) samples.

Brodwin and Muzzin likewise have used IRAC to construct stellar mass
selected samples of clusters beyond $z\simeq$1. Brodwin reports
a sample of over 116 systems whose redshifts likely lie beyond $z\simeq$1.
Together with progress in X-ray searches,  the route is now open
for reliable counts of clusters to $z\simeq$2.

\vspace{0.1in}
\centerline{\hbox{\psfig{file=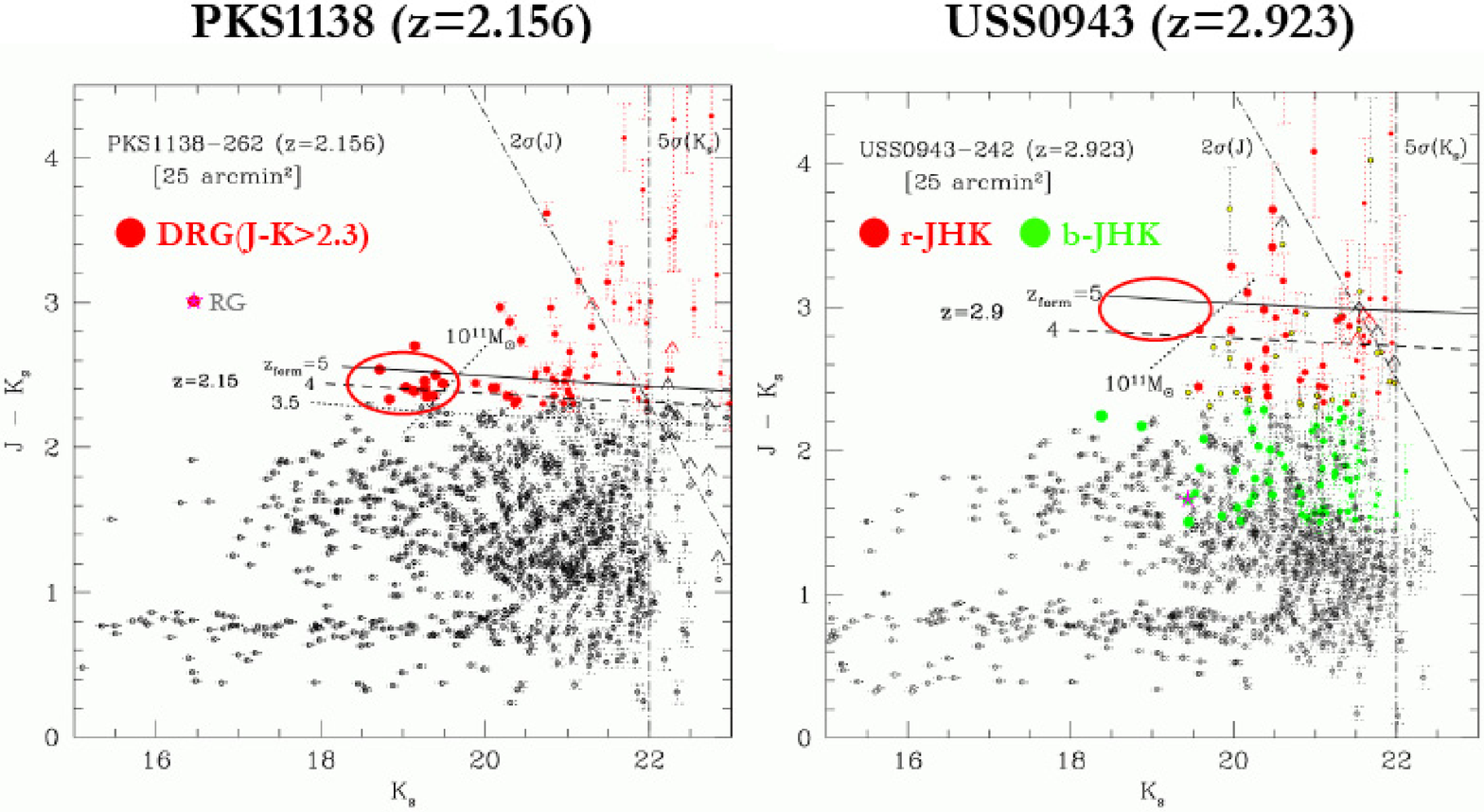,width=5.2in,angle=0}}}
\vspace{0.1in}
\noindent{\small \textsf{\bf Figure 7:} Infrared color-luminosity relation for
regions surrounding two forming clusters at $z$=2.15 (left) and $z$=2.9
(right) from the study of Kodama et al (2007). Highlighted sources represent
candidate members which constitute an emerging red sequence.} 
\smallskip

In addition to statistical studies, the importance of detailed multi-wavelength
studies of carefully-chosen individual systems has also been emphasized
by Moran, Okamura and Kurk (also in the associated posters by Koyama,
Nakata and Ma). A highlight of this work is the long-standing challenge
of identifying the precursors to the local cluster S0 population. The impact
of wide-field HST and GALEX imaging together with multi-object spectroscopy
was ably demonstrated by Moran who showed how there are sufficient passive spirals  -
morphological spirals with no ongoing star formation with excess UV
light consistent with recent activity -  to account for the local S0 population.

\section{Wide Field Galactic Archeology: Case for WFMOS}

The final session of the conference brought us close to home through
the use of imaging and spectroscopy to study our nearest neighbors.
Kodaira, Tamura, Freeman, Wyse, Guthakurta and Gilmore emphasized 
the revolution occurring in the dissection of both our Milky Way and M31 into
its various historic components. 

Galactic streams discovered in SDSS data (Belokurov et al 2006) and 
the intergalactic streams located between M31 and M33 (Guhathakurta,
Ibata et al 2007, Figure 8) contain information on past accretion events, the nature
of the dark matter halo and a close up view of the build up of Local Group 
galaxies that nicely complements the more uncertain high redshift probes.

This really is a growth area and one where truly panoramic imaging and
spectroscopic pays huge dividends. Colors of stars are a poor guide
to their pedigree and velocities and chemical measures are critical
to progress. This, of course, means spectroscopy! Much of the progress
we saw came from international groups who gained access to the DEIMOS
spectrograph on Keck.

Freeman and Wyse made a strong case for the Japanese community to embrace the
Wide Field Multi-Object Spectrograph (WFMOS) proposed as a sister
instrument to HyperSuprime-Cam at the Subaru prime focus. I'd like
to abuse my privelege as summary speaker to wholeheartedly support
their case. Time and again in astronomy we have seen the benefits of
a synergy between wide field imaging and spectroscopy, the 2dF and 
SDSS being the most recent examples. The Subaru community has
a fantastic (if admittedly expensive) opportunity to connect the now-funded
Hypersuprime-Cam imager with a state-of-the art multi-fiber spectrograph
matching precisely the same field. Even if all the scientific applications cannot
yet be envisaged, history tells us that this would be a terrific partnership
in instrumentation.

\vspace{0.1in}
\centerline{\hbox{\psfig{file=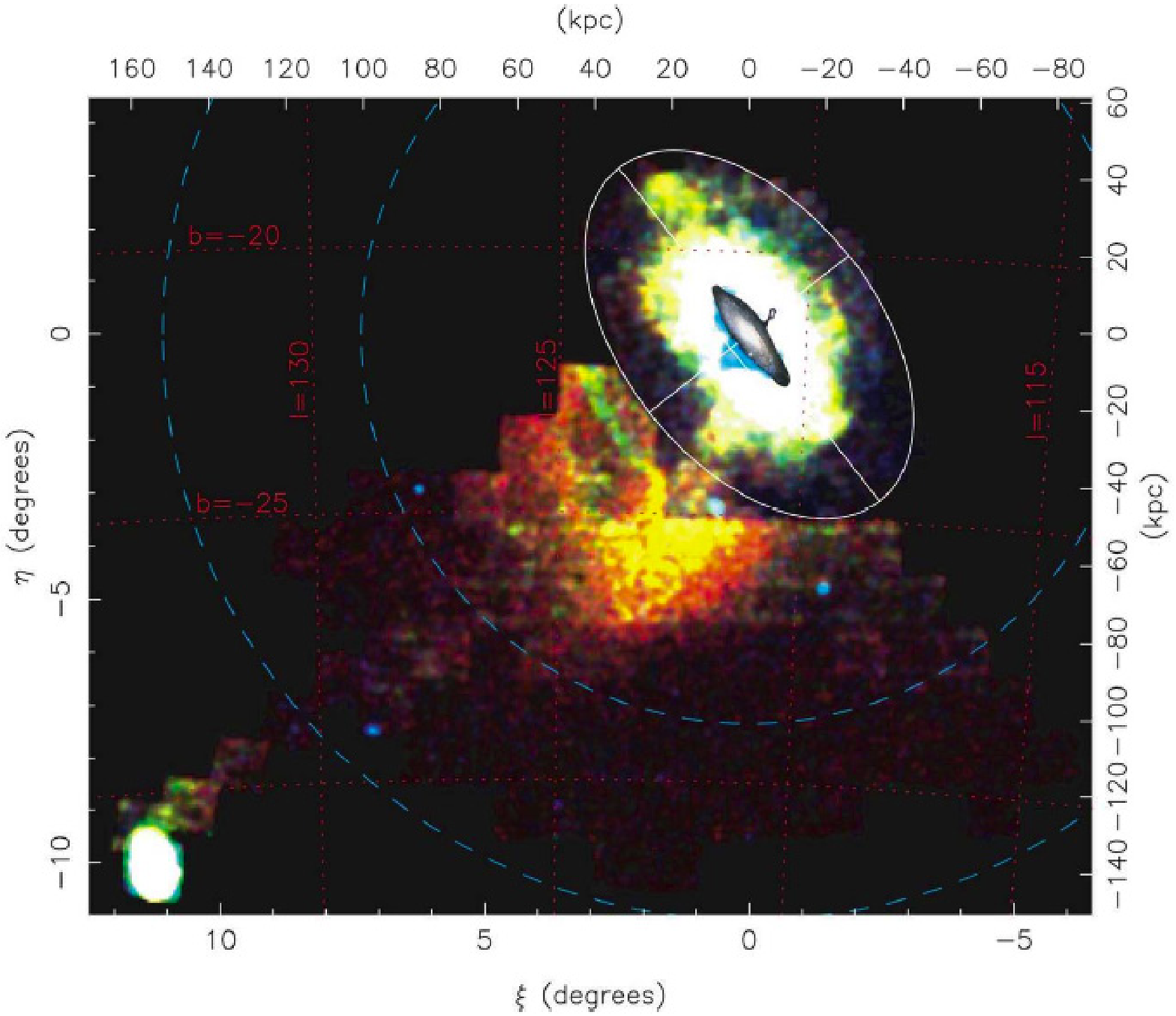,width=4.5in,angle=0}}}

\vspace{0.1in}
\noindent{\small \textsf{\bf Figure 8: } INT/CFHT survey of the southern
halo of M31; M33 can be seen in the lower left. The halos or M31 and
M33 overlap and are replete with substructure. The sheer angular scale 
of this image illustrates beautifully the potential of the 1.5 deg$^2$ field 
of Hypersuprime-Cam and its proposed associated WFMOS optical 
multi-fiber spectrograph.}


\acknowledgements 

I thank the lead organizers, Taddy Kodama and Toru Yamada for organizing
a terrific meeting and for their patience in waiting for this summary to be written.



\begin{thebibliography}{}
\bibitem[]{}Baugh, C et al 2005 MNRAS 356, 1191
\bibitem[]{}Belokurov, V et al 2006 Ap J 642, L137
\bibitem[]{}Bouwens, R et al 2007 Ap J 670, 928
\bibitem[]{}Bower, R G et al 2006 MNRAS 370, 645
\bibitem[]{}Bundy, K et al 2005 Ap J 625, 621
\bibitem[]{}Bundy, K et al 2006 Ap J 651, 120
\bibitem[]{}Bunker, A et al 2004 MNRAS 355, 374
\bibitem[]{}Croton, D et al 2006 MNRAS 365, 11
\bibitem[]{}Drory, N \& Alvarez, M 2008 Ap J 680, 41 
\bibitem[]{}Eyles, L et al 2005 MNRAS 364, 443
\bibitem[]{}Granato, G et al 2004 Ap J 600, 580
\bibitem[]{}Grothkopf, U et al 2007 ESO Messenger 128, 67
\bibitem[]{}Hayashi, M et al 2007 Ap J 660, 72
\bibitem[]{}Hirsch, J E, 2005 PNAS 102, 16569
\bibitem[]{}Hopkins, A \& Beacom, J 2006 Ap J 651, 142
\bibitem[]{}Ibata, R et al 2007 Ap J 671, 1591
\bibitem[]{}Iye, M et al 2006 Nature 443, 186
\bibitem[]{}Iye, M 2007 Highlights of Astronomy, 14, 532
\bibitem[]{}Kashikawa, N et al 2006
\bibitem[]{}Kodama, T et al 2007 MNRAS 377, 1717
\bibitem[]{}Massey, R J et al 2007 Nature 445, 286
\bibitem[]{}Okamoto, T et al 2008 MNRAS 385, 161
\bibitem[]{}Ouchi, M et al 2008 Ap J Supp 176, 301
\bibitem[]{}Rauch, M et al 2007 astro-ph/0711.1354
\bibitem[]{}Ryan-Weber, E et al 2006 MNRAS 371, L78
\bibitem[]{}Smail, I et al 2007 Ap J 654, L33
\bibitem[]{}Springel, V et al 2005a MNRAS 361, 776
\bibitem[]{}Springel, V et al 2005b Nature 435, 629
\bibitem[]{}Stark, D et al 2007 Ap J 659, 84
\bibitem[]{}Stark, D et al 2008 Nature, in press
\bibitem[]{}Stringer, M. et al 2008 astro-ph/0806.2232
\bibitem[]{}Swinbank, M et al 2006 MNRAS 368, 1631
\bibitem[]{}Wilkins, S et al 2008 MNRAS 385, 687

\end{thebibliography}
\end{document}